\pdfoutput=1

\documentclass[twocolumn]{aastex63}
\usepackage{natbib}
\usepackage{hyperref}
\usepackage{lineno}
\usepackage{ulem}

\def\deg{\ifmmode^\circ\else$^\circ$\fi}

\shorttitle{Probing physical processes in G47 filamentary cloud}
\shortauthors{O.~R. Jadhav et al.}

\begin{document}
\title{Unveiling Physical Conditions and Star Formation Processes in the G47 Filamentary Cloud}
\correspondingauthor{O.~R. Jadhav}
\email{Email: omkar@prl.res.in}

\author[0009-0001-2896-1896]{O.~R. Jadhav}
\affiliation{Astronomy \& Astrophysics Division, Physical Research Laboratory, Navrangpura, Ahmedabad 380009, India}
\affiliation{Indian Institute of Technology Gandhinagar Palaj, Gandhinagar 382355, India}

\author[0000-0001-6725-0483]{L.~K. Dewangan}
\affiliation{Astronomy \& Astrophysics Division, Physical Research Laboratory, Navrangpura, Ahmedabad 380009, India}

\author[0000-0003-4941-5154]{A. Haj Ismail}
\affiliation{College of Humanities and Sciences, Ajman University, 346 Ajman, United Arab Emirates}

\author[0000-0001-8812-8460]{N.~K.~Bhadari}
\affiliation{Astronomy \& Astrophysics Division, Physical Research Laboratory, Navrangpura, Ahmedabad 380009, India}

\author[0000-0002-7367-9355]{A.~K. Maity}
\affiliation{Astronomy \& Astrophysics Division, Physical Research Laboratory, Navrangpura, Ahmedabad 380009, India}
\affiliation{Indian Institute of Technology Gandhinagar Palaj, Gandhinagar 382355, India}

\author[0000-0002-6740-7425]{Ram Kesh Yadav}
\affiliation{National Astronomical Research Institute of Thailand (Public Organization), 260 Moo 4, T. Donkaew, A. Maerim, Chiangmai 50180, Thailand}

\author[0000-0002-2726-1388]{Moustafa Salouci}
\affiliation{Strategic Planning and Institutional Identity Administration, King Faisal University, Al-Ahsaa 31982, Saudi Arabia}

\author[0000-0002-7125-7685]{Patricio Sanhueza}
\affiliation{Department of Astronomy, School of Science, The University of Tokyo, 7-3-1 Hongo, Bunkyo, Tokyo 113-0033, Japan}
\affiliation{National Astronomical Observatory of Japan, National Institutes of Natural Sciences, 2-21-1 Osawa, Mitaka, Tokyo 181-8588, Japan.}

\author[0000-0001-5731-3057]{Saurabh Sharma}
\affiliation{Aryabhatta Research Institute of Observational Sciences, Manora Peak, Nainital 263002, India}

\begin{abstract}
We present a multi-wavelength study of the filamentary cloud G47 (d $\sim$4.44 kpc), 
which hosts the mid-infrared bubbles N98, B1, and B2. 
The SMGPS 1.3 GHz continuum map detects ionized emission toward all the bubbles, marking the first detection of 
ionized emission toward the B2 bubble. Analysis of the unWISE 12.0 $\mu$m image, {\it Spitzer} 8.0 $\mu$m image, and the {\it Herschel} column density 
and temperature maps reveals two previously unreported hub-filament system candidates associated with the H\,{\sc ii} regions B2 and N98, 
which are powered by massive OB stars. This indirectly favours the applicability of a global non-isotropic collapse (GNIC) scenario for 
massive star formation in N98 and B2. 
The position-position-velocity diagram of FUGIN $^{13}$CO(1--0) shows significant velocity variations from 61 to 53 km s$^{-1}$ toward 
areas between B2 and N98, where the  magnetic field morphology exhibits significant curvature, and high velocity dispersion (i.e., 2.3--3.1 km s$^{-1}$) 
is observed. This may be explained by the expansion of the H\,{\sc ii} regions B2 and N98. The energy budget of the cloud, estimated using 
SOFIA/HAWC+ and molecular line data, suggests that the magnetic field dominates over turbulence and gravity in G47. Furthermore, the radial column 
density and velocity profiles of G47 display signatures of converging flows in a sheet-like structure. The relative orientations between the 
magnetic field and local gravity suggest that G47 may undergo gravitational contraction along the magnetic field lines once it becomes 
magnetically supercritical. 
\end{abstract}

\section{Introduction}
Infrared observations from space-based telescopes, such as {\it Spitzer} 
and {\it Hershel}, have unveiled the presence of mid-infrared (MIR) bubbles and filamentary structures in massive star-forming regions \citep{Churchwell_2006, Andre_2014}. 
Filaments, in particular, have gained significant attention in star formation studies, as they are now recognized as central structures playing a pivotal role in the star formation processes. These filaments often arrange themselves into hub-filament systems \citep[HFSs;][]{myers09}, which are fundamental structures driving the birth of massive OB stars ($\ge$ 8 M$_{\odot}$) and the development of star clusters \citep[e.g.,][]{Kumar_2020, liu23, yang23}.

On the other hand, MIR bubbles, often associated with massive stars, are frequently seen embedded within filaments. The radiative and mechanical energy, emitted by these massive stars, significantly influences and shapes their surrounding environments. Such systems offer valuable opportunities to study the escape and trapping of ionizing radiation emitted by an O-type star forming within a filament \citep[e.g.,][]{whitworth21}. 

The impact of expanding H\,{\sc ii} regions on the filaments is governed by several factors, such as the orientation of the filament with respect to the ionizing source, the density and temperature of the filament, and the strength and configuration of magnetic fields (B-fields). \citet{Arthur_2011} examined the role of B-fields in the evolution of H\,{\sc ii} regions and concluded that B-fields may inhibit the fragmentation of neutral gas swept up by 
expansion of these H\,{\sc ii} regions. Ultimately, the evolution of filaments affected by H\,{\sc ii} regions is determined by the complex interplay of gravity, turbulence, and B-fields \citep{Federath_2016}. Nevertheless, the origins and feedback processes of massive stars are still not well constrained and remain active areas of study \citep{zinnecker07,tan14,Motte+2018,rosen20}.
 
\begin{table*}
\scriptsize
\setlength{\tabcolsep}{0.1in}
\centering
\caption{List of multi-wavelength surveys used in this paper.}
\begin{tabular}{lcccr}
\hline
\hline
 Survey  &  wavelength/ line(s)      &  Angular Resolution ($''$)        &  Reference \\   
\hline
SMGPS &1.3 GHz& $\sim$8& \citet{Goedhart_2024}\\


FUGIN
& $^{13}$CO(1--0) & $\sim$20       &\citet{10.1093/pasj/psx061}\\

ATLASGAL & 870 $\mu$m & $\sim$19 &\citet{Schuller_2009}\\

Hi-GAL       &70, 160, 250, 350, 500 $\mu$m                     &$\sim$6, 12, 18, 25, 37   &\citet{Molinari}\\

FIELDMAPS
&214 $\mu$m &$\sim$18    &\citet{Stephens_2022}\\

MIPSGAL	&24 $\mu$m 	&$\sim$6	&\citet{Carey_2009}\\ 

unWISE Survey &12 $\mu$m &$\sim$6 &\citep{Lang_2014}\\

GLIMPSE      &3.6, 4.5, 5.8, 8.0 $\mu$m                   & $\sim$2          &\citet{Benjamin_2003}\\

\hline         
\end{tabular}
\label{data_tab}
\end{table*}

 
This paper focuses on the G47 filamentary cloud (hereafter G47 cloud), situated at a distance ($d$) of $\sim$4.44 kpc \citep{10.1093/mnras/stv735}, in the close vicinity of the Sagittarius Far Arm of the Milky way galaxy \citep{Zucker_2018}. G47 cloud is a good representative of a class of objects called Infrared Dark Clouds (IRDCs), which have been suspected as sites of massive star formation for a while \citep{Rathborne_2006,Chambers_2009,Sanhueza_2012,Goodman_2014,Sanhueza_2019}. Previously, four MIR bubbles---B1, B2, N97, and N98 \citep[see also][]{Churchwell_2006}---were reported toward the G47 cloud \citep{Xu_2018}, which hosts massive stars and young stellar objects (YSOs). The locations of the MIR bubbles are indicated in the {\it Spitzer} color composite map (see Figure~\ref{fg1}a). In Figure~\ref{fg1}b, the filamentary appearance of the G47 cloud is clearly depicted in the {\it Herschel}  color composite map. \citet{Xu_2018} found that turbulence plays a dominant role over gravity in the G47 cloud. In general, the plane-of-the-sky (POS) B-field structure in the ISM can be traced using the polarimetric observations \citep{Crutcher_2012, Andersson_2015}. Using the SOFIA 214 $\mu$m emission polarization data, the POS B-field structure toward the G47 cloud was studied by \citet{Stephens_2022}. They found that in the densest, active star-forming regions of the cloud, the B-fields are predominantly aligned perpendicular to the filament's major axis. In contrast, in other areas, the B-fields show either a parallel or more random orientation, resulting in an overall curved morphology.

Despite existing studies on the G47 cloud, a detailed investigation of the MIR bubbles powered by massive stars, including an in-depth analysis of the B-field configuration and gas kinematics remains absent. Furthermore, the energy budget of the cloud, simultaneously encompassing gravitational, kinetic, and magnetic energies, 
has yet to be determined. 
Moreover, previous studies in the literature have investigated some, but not all, of the physical processes driving the formation of massive stars in the G47 cloud \citep[e.g.,][]{Xu_2018,Stephens_2022}.

This study employs a multi-wavelength approach to investigate the physical environment and star formation in the G47 cloud, with a focus on the MIR bubbles in G47. We carried out a detailed kinematic analysis of embedded structures using the $^{13}$CO(1--0) line data from the FOur-beam REceiver System on the 45 m Telescope (FOREST) Unbiased Galactic plane Imaging survey with the Nobeyama 45-m telescope \citep[FUGIN;][]{Minamidani_2016,10.1093/pasj/psx061}. We also analyzed the publicly available 1.3 GHz continuum map from the South African Radio Astronomy Observatory (SARAO) MeerKAT Galactic Plane Survey \citep[SMGPS;][]{Goedhart_2024}. In addition, we revisited the published SOFIA 214 $\mu$m emission polarization data \citep{Stephens_2022} to explore the B-field morphology associated with the MIR bubbles, and performed an energy budget analysis for the G47 cloud.

Section~\ref{data} presents the observational data sets analyzed in this paper. The results are provided in Section~\ref{results}. Section~\ref{discussion} discusses the interpretations of our findings. Finally, the conclusions of this study are given in Section~\ref{conclusion}.
\section{Observational datasets and analysis}
\label{data}
In this work, we analyzed publicly available multi-wavelength data sets toward the G47 cloud, primarily in the area of 0\rlap.{$^\circ$}43 $\times$ 0\rlap.{$^\circ$}72
 centered at Galactic longitude (\textit{l}) = 47\rlap.{$^\circ$}05 and Galactic latitude (\textit{b}) = 0\rlap.{$^\circ$}31. An overview of the data sets used in this
work is listed in Table~\ref{data_tab}. A summary of all the data sets analyzed in this study is presented below.

\subsection{NIR and MIR data}
We acquired NIR and MIR images at 3.6--8.0 $\mu$m (resolution $\sim$2$''$; plate scale $\sim$0\rlap.{$''$}6) and photometric magnitudes of point-like sources in these wavelengths from the {\it Spitzer} Galactic Legacy Infrared Mid-Plane Survey Extraordinaire \citep[GLIMPSE;][]{Benjamin_2003} survey. The {\it Spitzer} Multiband Imaging Photometer for Spitzer \citep[MIPS;][]{Rieke_2004} image at 24.0 $\mu$m (resolution $\sim$6$''$; plate scale $\sim$1\rlap.{$''$}25) was downloaded from the MIPS Inner Galactic Plane Survey \citep[MIPSGAL;][]{Carey_2009}.  Additionally, the MIR 12.0 $\mu$m image was collected from the Unblurred Coadds of the Wide-field Infrared Survey Explorer (WISE) Imaging (unWISE) survey \citep{Lang_2014}, which provides WISE images with an improved signal-to-noise ratio.
\subsection{Dust continuum data}
\subsubsection{ SOFIA data}
\label{ss2.5}
The G47 filamentary cloud was observed with the SOFIA telescope using the High-resolution Airborne Wideband Camera Plus (HAWC+) instrument \citep{Harper_2018} at 214 $\mu$m (or BAND E). These observations were taken as a part of the FIELDMAPS Legacy project  \citep[PI: Ian Stepehens; Proposal ID: 08$\_$0186; see also][]{Stephens_2022}. The final data products have a spatial resolution of $\sim$18\rlap.{$''$}2 and a pixel scale of $\sim$4\rlap.{$''$}55.
\subsubsection{{\it Herschel} data}
\label{2.5}
The processed images at 70 $\mu$m, 160 $\mu$m, 250 $\mu$m, 350 $\mu$m, and
500 $\mu$m were collected from the {\it Herschel} Space Observatory data archives, which were taken as a part of {\it Herschel} infrared Galactic Plane Survey \citep[Hi-GAL;][]{Molinari}. The 70 and 160 $\mu$m images were obtained with the Photodetector Array Camera and Spectrometer \citep[PACS;][]{Poglitsch_2010}, 
while the images at 250, 350, and 500 $\mu$m were acquired with the SPectral and photometric Imaging REceiver \citep[SPIRE;][]{Griffin_2010}. 
The resolutions of these images are $\sim$5\rlap.{$''$}8, 12$''$, 18$''$, 25$''$, and 37$''$ \citep{Poglitsch_2010,Griffin_2010}, respectively.

The \textit{Herschel} images at 70--500 $\mu$m allow us to derive the column density ($N(\rm H_2$)) and the dust temperature ($T_{\rm d}$) maps. These maps can be generated by performing pixel-by-pixel spectral energy distribution (SED) fitting of the \textit{Herschel} multi-wavelength flux using a modified black body equation \citep[e.g.,][]{2023MNRAS.523.5388M, 2023ApJ...946...22D, Dewangan_2024a}. The modified black body equation is given by \citep{Battersby_2011}
\begin{equation}
S_{\nu} = \frac{2h\nu^{3}}{c^{2} (e^{\frac{h\nu}{kT}} - 1)}
(1 - e^{-\tau_{\nu}}),
\end{equation}
where 
\begin{equation}
\tau_{\nu} = \mu_{\rm H_{2}}  m_{\rm H} \kappa_{\nu} N(\rm H_{2}), 
\end{equation}
where $\mu_{\rm H_{2}}$ is the mean molecular weight, $m_{\rm H}$ is the mass of
hydrogen, and $\kappa_{\nu}$ is the dust
opacity.
However, for this pixel-by-pixel fitting, all \textit{Herschel} images must be convolved to the same resolution and pixel scale. This method therefore produces the $N(\rm H_2$) and $T_{\rm d}$ maps with a resolution of $\sim$36\rlap.{$''$}3. Considering this limitation, we employed \texttt{HIRES} \citep{Menschikov} to generate the high resolution $N(\rm H_2$) and $T_{\rm d}$ maps, which provides resolutions ranging from $\sim$8$''$ to $\sim$36$''$.  
The basic concept behind these high-resolution maps involves adding higher-resolution details to the lower-resolution images as differential terms \citep[see Equation A.1 in][]{Palmeirim_2013} \citep[Equation 5 in][]{Menschikov}.
The \texttt{HIRES} fitting procedure assumes that (1) the original images represent optically thin thermal emission from dust grains with a power-law opacity $\kappa_{ \lambda} \propto \lambda^{-\beta}$ and a constant value of spectral emissivity ($\beta$=2), (2) the dust temperature is constant along the lines of sight passing through
each pixel of the images, and (3) the lines of sight are free from contamination by unrelated radiation, both in front of and behind the observed structures \citep[see more details in][]{Menschikov}.
The resulting $N(\rm H_2$) and $T_{\rm d}$ maps generated using \texttt{HIRES} at 8$''$ resolution, were found to be noisy due to the presence of heated dust in the 70 $\mu$m image. Consequently, the final maps with a resolution of $\sim$13\rlap.{$''$}5 are used in this work (see Section~\ref{HIRES}).
\subsubsection{Sub-millimeter 870 $\mu$m data}
\label{aa2.5}
The 870 $\mu$m emission data (resolution $\sim$18\rlap.{$''$}2)  were sourced from the Atacama Pathfinder Experiment (APEX) Telescope Large Area Survey of the GALaxy  \citep[ATLASGAL;][]{Schuller_2009}. 
The catalog by \citet{Urquhart_2017} details about 8000 dense clumps in the Galactic disc using the ATLASGAL data. They also provided the velocities, distances, luminosities, and masses of clumps \citep[see][for more details]{Urquhart_2017}. Using this catalog, we identified the positions of dense clumps associated with the G47 cloud.
\subsection{Molecular line Data}
In the direction of G47, we conducted a detailed analysis of the FUGIN $^{13}$CO(1--0) line data \citep[rms $\sim$0.7 K; spatial resolution $\sim$20\rlap.{$''$}2; pixel scale $\sim$8\rlap.{$''$}5; velocity resolution $\sim$1.3 km s$^{-1}$;][]{Minamidani_2016,10.1093/pasj/psx061}, which were calibrated in main beam temperature ($T_{\rm mb}$). To enhance the signal-to-noise ratio, the FUGIN line data were smoothed using a Gaussian function with a 1 pixel half-power beamwidth, resulting in a resolution of $\sim$22$''$.
\subsection{Radio Data}
The radio continuum emission map at 1.3 GHz from the SMGPS \citep{Goedhart_2024} was explored in the direction of G47. The data offer an angular resolution of 
$\sim$8\rlap.{$''$}0 and a  sensitivity of $\sim$10--20 $\mu$Jy beam$^{-1}$.
\section{Results}
\label{results}
\subsection{Physical environment in the G47 cloud}
\label{HIRES}
\begin{figure*}
\centering
\includegraphics[width=15 cm]{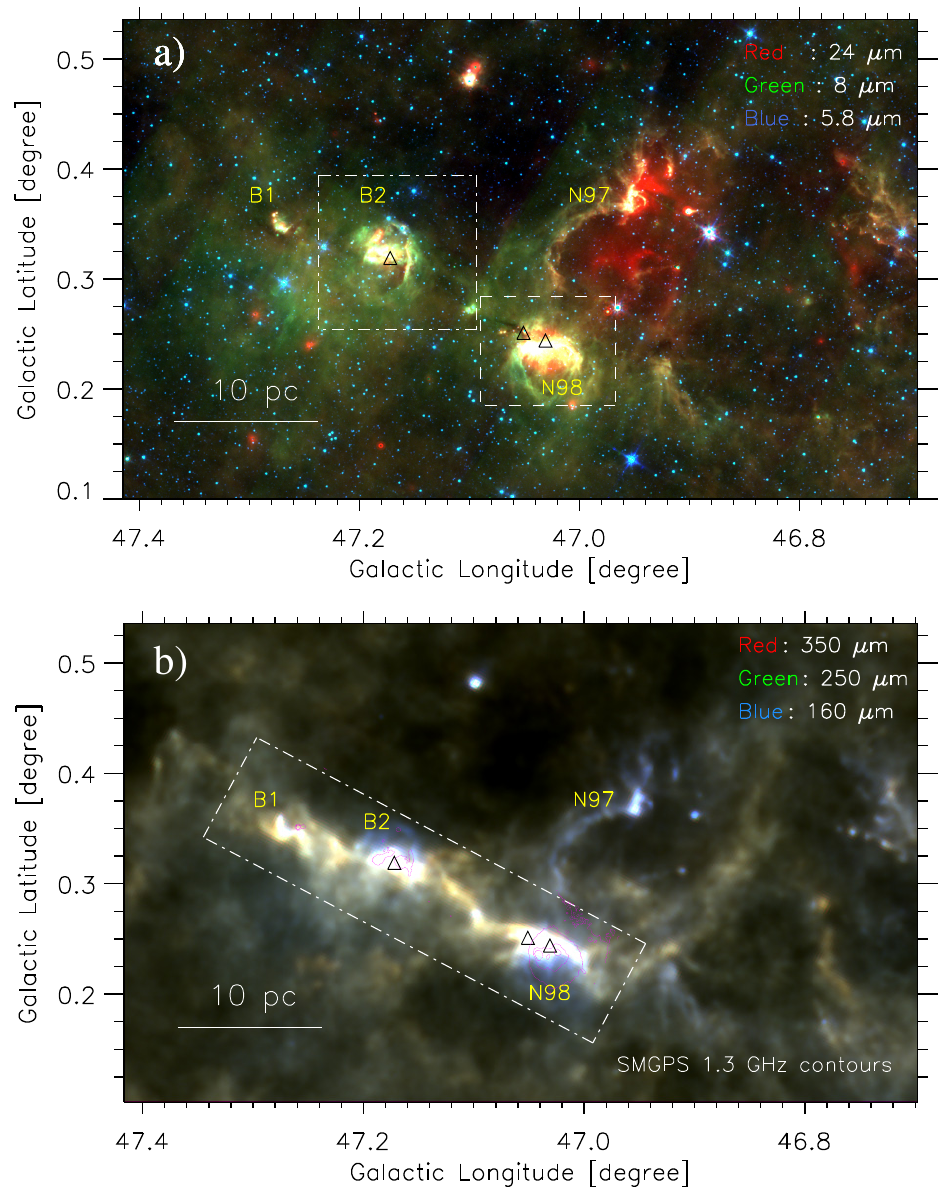}
\caption{a) \textit{Spitzer} three-color composite map (red: 24 $\mu$m, green: 8 $\mu$m, blue: 5.8 $\mu$m) of a larger area hosting the MIR bubbles (size $\sim$0\rlap.{$^\circ$}43 $\times$ 0\rlap.{$^\circ$}72; cental coordinates $\sim$(\textit{l} = 47\rlap.{$^\circ$}05; \textit{b} = 0\rlap.{$^\circ$}31)). The region enclosed by the dot-dashed box corresponds to the area displayed in Figure~\ref{fg3}a, while the dashed box outlines the area shown in Figures~\ref{fg4}a--\ref{fg4}d.
b) \textit{Herschel} three-color composite map (red: 350 $\mu$m, green: 250 $\mu$m, and blue: 160 $\mu$m). The dot-dashed box encompasses the area shown in Figures~\ref{fg2}a and~\ref{fg2}b. The SMGPS 1.3 GHz radio continuum contours (in magenta) are also shown toward the area highlighted by the dot-dashed box, and the contour levels are 0.25, 2, and 3 mJy beam$^{-1}$. 
The locations of previously known MIR bubbles N98, N97, B1, and B2 are indicated in both the panels. 
In each panel, the open triangles highlight the positions of the ATLASGAL clumps, and the scale bar corresponds to 10 pc derived at a distance of 4.44 kpc.}
\label{fg1} 
\end{figure*}

In the direction of the G47 cloud, the spatial morphology of the MIR bubbles (i.e., N97, N98, B1, and B2) and the filamentary structure can be examined in Figures~\ref{fg1}a and~\ref{fg1}b. 
Figure~\ref{fg1}a presents a three-color composite map produced from {\it Spitzer} 24 $\mu$m (in red), 8 $\mu$m (in green), and 5.8 $\mu$m (in blue) images, while Figure~\ref{fg1}b shows a three-color composite map generated from \textit{Herschel} 350 $\mu$m (in red), 250 $\mu$m (in green), and 160 $\mu$m (in blue) images. 
The {\it Herschel} color composite map is also overlaid with the SMGPS 1.3 GHz radio continuum emission contours. The NRAO VLA Sky Survey radio continuum emission at 1.4 GHz (resolution $\sim$45$''$, sensitivity $\sim$0.45 mJy beam$^{-1}$) was detected toward B1, N97, and N98, but no radio continuum emission was detected toward the B2 bubble \citep{Xu_2018}. 
One can notice that the SMGPS radio continuum map at 1.3 GHz reveals emission from all the bubbles, including the B2 bubble (see magenta contours in Figure~\ref{fg1}b). Due to the improved sensitivity \citep[root-mean-square sensitivity $\sim$10--20 $\mu$Jy beam$^{-1}$;][]{Goedhart_2024} of the SMGPS data, faint radio emission toward the B2 bubble has been detected for the first time.

\begin{figure*}
\centering
\includegraphics[width=15cm]{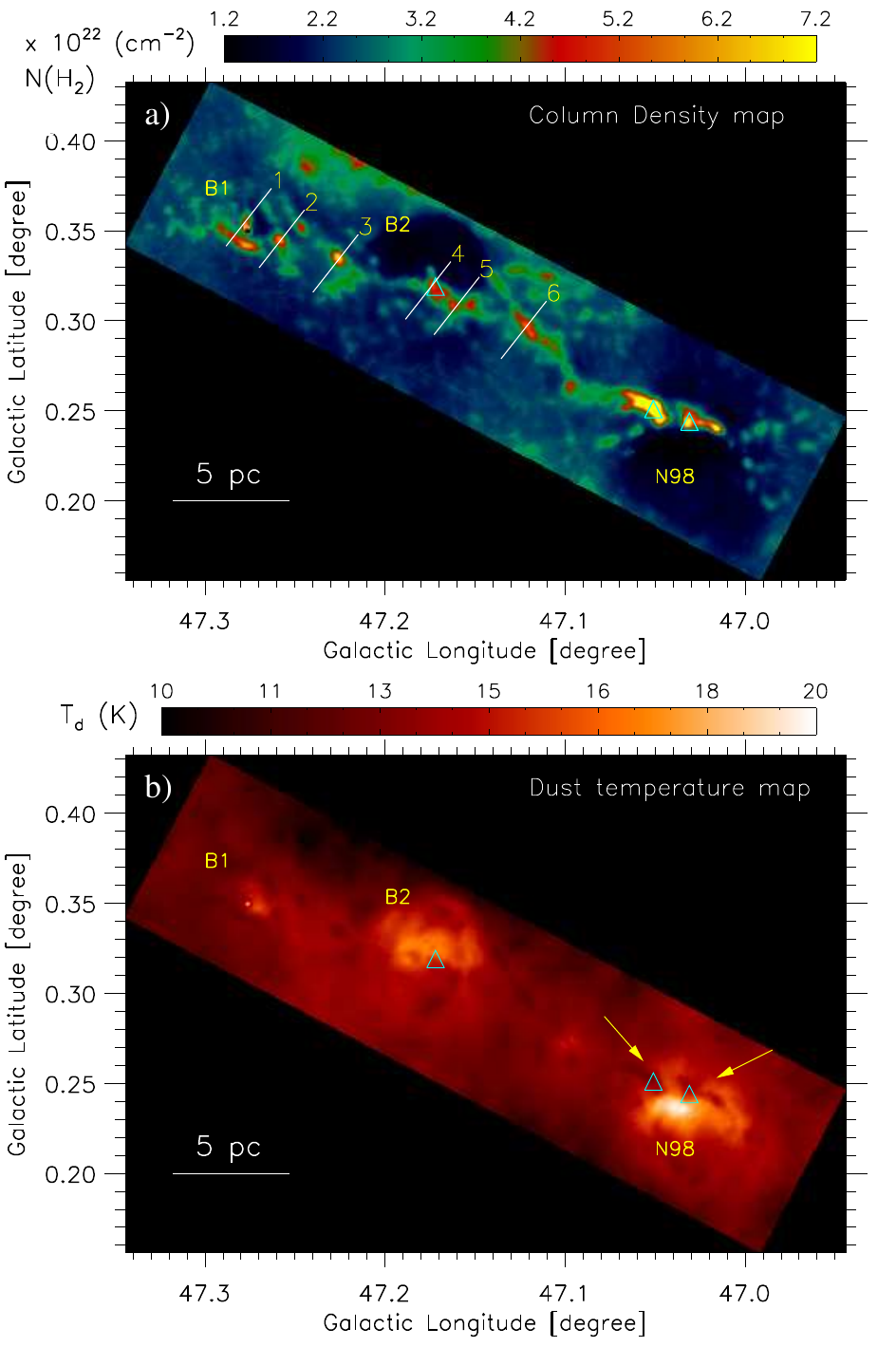}
\caption{a) \textit{Herschel} column density map overlaid with six white lines, where the velocity and column density profiles are extracted (see Figure~\ref{fg11}). b) \textit{Herschel} dust temperature map. Both images (resolution $\sim$13\rlap.{$''$}5) are generated using the \texttt{HIRES} tool (see section~\ref{2.5}). 
In each panel, the scale bar corresponds to 5 pc derived at a distance of 4.44 kpc, and the symbols are same as Figure~\ref{fg1}a.} 
\label{fg2}
\end{figure*}

The {\it Herschel} $N(\rm H_2$) and $T_{\rm d}$ maps of the G47 cloud are shown in Figures~\ref{fg2}a and~\ref{fg2}b, respectively (see the dot-dashed box in Figure~\ref{fg1}b). 
The positions of the ATLASGAL clumps are also marked in Figures~\ref{fg2}a and~\ref{fg2}b. 
Figure~\ref{fg2}a displays the presence of several condensations with higher values of $N(\rm H_2)$ (i.e., $\sim$4--7 $\times$ 10$^{22}$ cm$^{-2}$). The presence of warm dust emission ($T_{\rm d}$ $\sim$17--20 K)  is found toward the B1, B2, and N98 bubbles in the $T_{\rm d}$ map (see Figure~\ref{fg2}b), along with a relatively low $N(\rm H_2$) ($<$ 10$^{22}$ cm$^{-2}$). Additionally, a dark feature with $T_{\rm d}$ $\sim$10 K is traced near the N98 bubble (see arrows in Figure~\ref{fg2}b), and appears to be well correlated with higher $N(\rm H_2)$ regions.

\begin{figure*}
 \centering
 \includegraphics[width=15 cm]{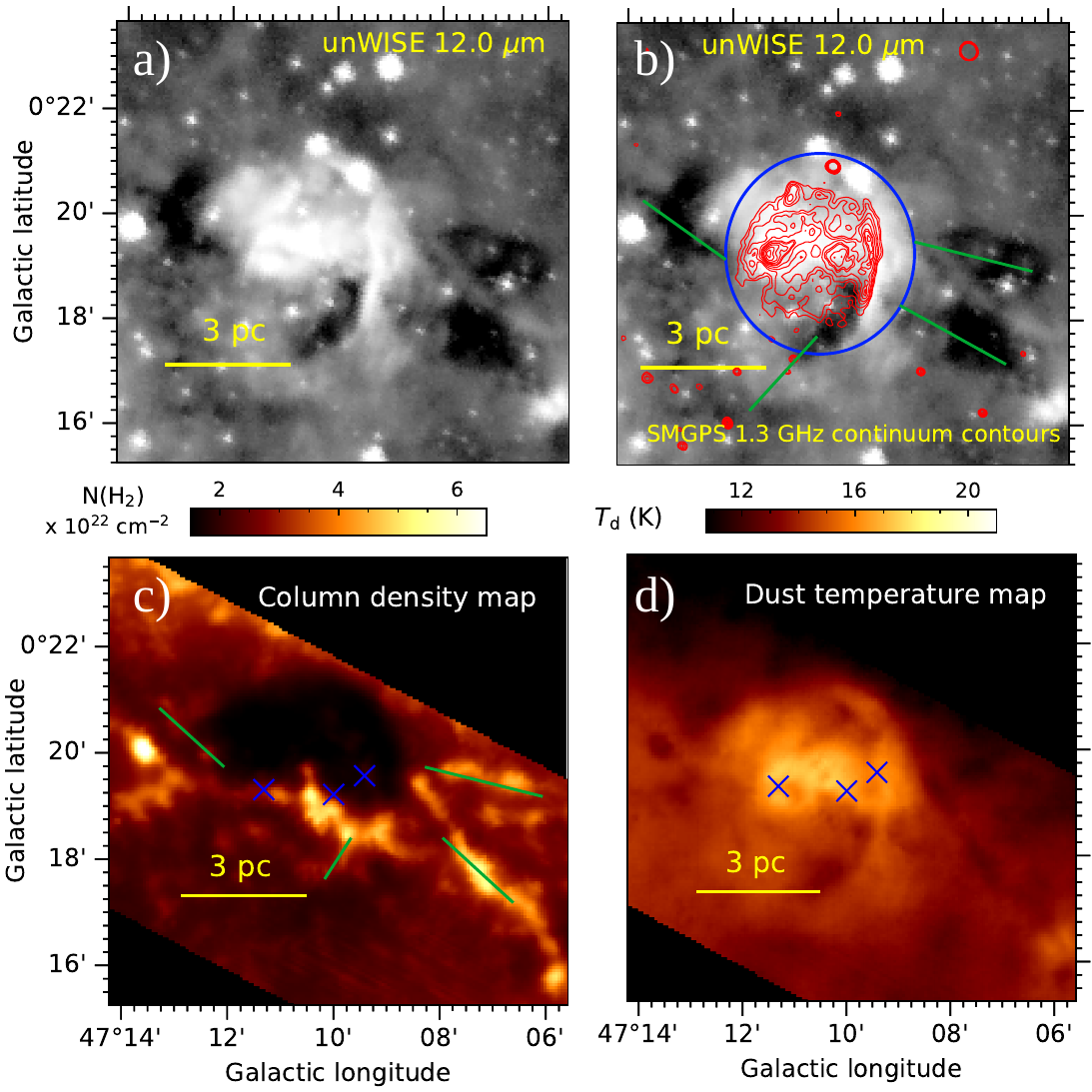}
\caption{a) The panel displays the unWISE 12.0 $\mu$m image (in the inverse hyperbolic sine scale) toward the B2 bubble 
(see the dot-dashed box in Figure~\ref{fg1}a). b) Same as Figure~\ref{fg3}a, but overaid with the SMGPS 1.3 GHz continuum emission contours. The contour levels are 0.11, 0.13, 0.17, 0.2, 0.25, 0.3, 0.32, 0.36, 0.4, 0.45, and 0.50 mJy beam$^{-1}$. A HFS candidate is indicated by a solid blue circle and green lines (see Section~\ref{HIRES} for more details).
c) {\it Herschel} column density map toward the B2 bubble. The solid green lines highlight the visually identified filamentary structures. 
d) {\it Herschel} dust temperature map the B2 bubble. 
In panels ``c" and ``d", the `$\times$' symbols mark the peak positions of the radio continuum emission toward the B2 bubble. 
The scale bar derived at a distance of 4.44 kpc is shown in all the panels.}
 \label{fg3}
\end{figure*}

\begin{figure*}
 \centering
 \includegraphics[width=15 cm]{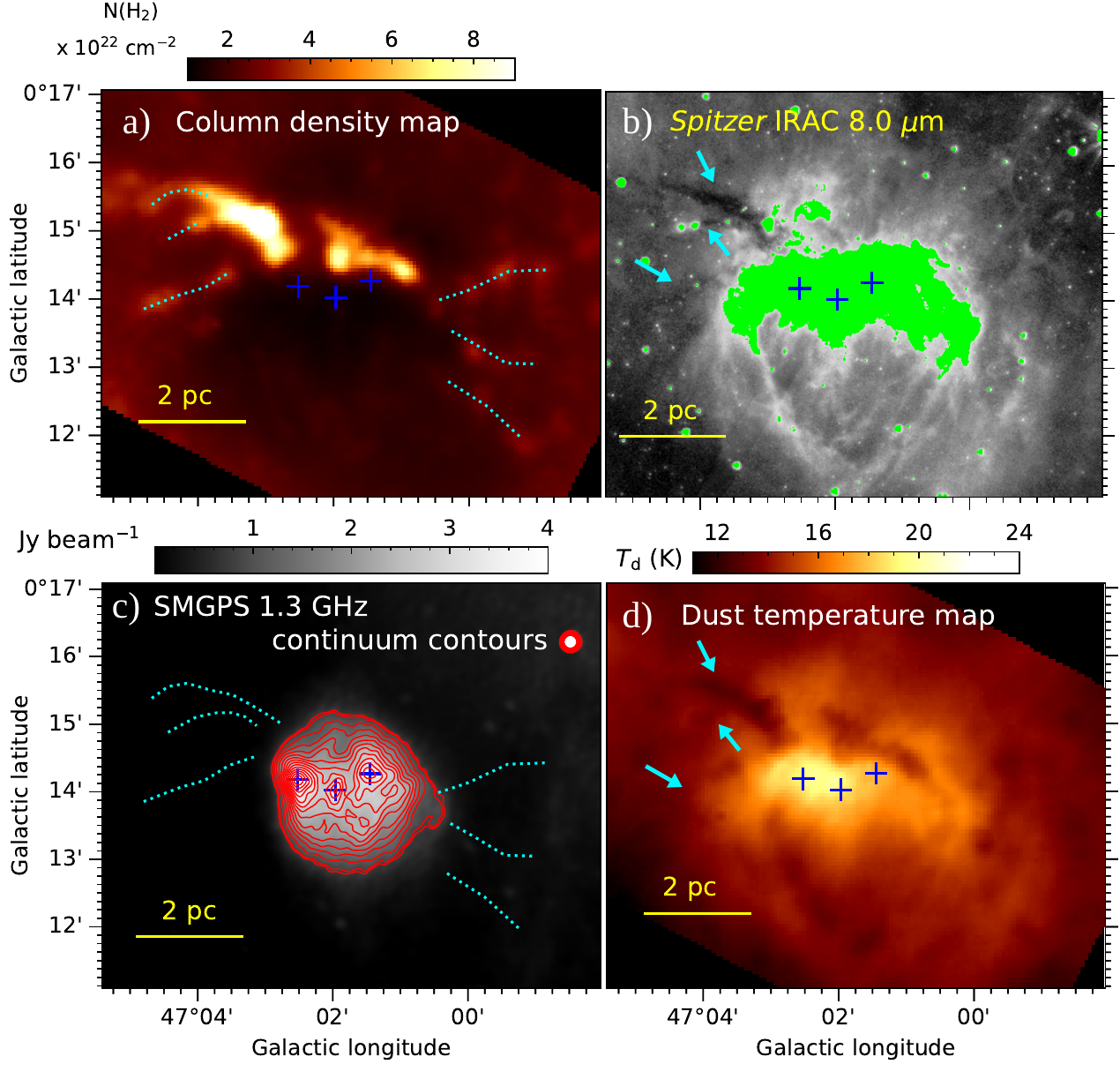}
 \caption{a) {\it Herschel} column density map toward N98 bubble (see the dashed box in Figure~\ref{fg1}a). The dotted lines highlight the filamentary structures. b) The panel displays the {\it Spitzer} 8.0 $\mu$m image (in the inverse hyperbolic sine scale) of the N98 bubble, highlighting its structural features with enhanced contrast for better visualization of both bright and faint regions. The filled contour level (in green) at 52 MJy sr$^{-1}$ is used to enhance the visibility of the absorption features. c) Overlay of SMGPS 1.3 GHz continuum contours on its corresponding emission map. The contour levels are 0.74, 0.82, 1.07, 1.32, 1.58, 1.83, 2.09, 2.34, 2.60, 2.85, 3.11, 3.36, 3.62, 3.87, 4.13, and 4.38 mJy beam$^{-1}$. d) {\it Herschel} dust temperature map. In each panel, the `$+$' symbols mark the peak positions of the radio continuum emission toward the N98 bubble, and the scale bar corresponds to 2 pc derived at a distance of 4.44 kpc.}
 \label{fg4}
\end{figure*}

To investigate the embedded structures, zoomed-in views of the B2 and N98 bubbles using multi-wavelength data are presented in Figures~\ref{fg3} and~\ref{fg4}, respectively. 
In Figure~\ref{fg3}a, we display the unWISE 12.0 $\mu$m image of the area hosting B2, where the ATLASGAL clump is located (see Figure~\ref{fg2}a). 
Figure~\ref{fg3}a reveals the bubble-like morphology (in emission) surrounded by absorption structures. 
The unWISE 12.0 $\mu$m image is also overlaid with the SMGPS 1.3 GHz continuum emission contours, highlighting the presence of substructures and multiple peaks in the ionized emission (see Figure~\ref{fg3}b). 
Considering the locations of the bubble-like morphology, absorption structures, and radio continuum emission, in Figure~\ref{fg3}b, a hub-filament configuration is highlighted by a solid circle and green lines. 
In the $N(\rm H_2$) map, visual inspection reveals at least four filamentary structures (see solid lines in Figure~\ref{fg3}c) surrounding the central 
clump (or bubble-like morphology), and these filaments are also seen as absorption structures in the unWISE 12.0 $\mu$m image. 
This morphology suggests the presence of a HFS \citep{myers09,Morales_2019} candidate toward the B2 bubble. 
Figure~\ref{fg3}d shows the distribution of warm dust emission toward B2, where the radio continuum emission is present.

A zoomed-in view of the N98 bubble is presented in Figures~\ref{fg4}a,~\ref{fg4}b,~\ref{fg4}c, 
and~\ref{fg4}d, using the $N(\rm H_2$) map, the {\it Spitzer} 8.0 $\mu$m image, the SMGPS 1.3 GHz continuum 
emission, and $T_{\rm d}$ map, respectively. The {\it Spitzer} 8.0 $\mu$m image displays a bubble-like morphology 
in emission, with its central area appearing to host both the radio continuum and warm dust emissions. 
Additionally, some absorption structures are also seen in the {\it Spitzer} 8.0 $\mu$m image, 
where cold dust emisison is detected just outside the bubble-like morphology. 
In order to highlight the absorption structures against the higher intensity regions in the {\it Spitzer} 8.0 $\mu$m image, 
a filled green contour is shown in Figure~\ref{fg4}b. These absorption structures also appear as emission features in the $N(\rm H_2)$ map, 
and can be interpreted as filaments. These filaments surrounding the ionized emission are indicated in Figure~\ref{fg4}c. 
This suggests the presence of a HFS candidate in the $N(\rm H_2)$ map toward the N98 bubble (see Figure~\ref{fg4}a).
\subsection{FUGIN $^{13}$CO(1--0) moment maps}
\label{molgas}
\citet{Xu_2018} studied the gas kinematics of the G47 cloud using $^{13}$CO(1--0) line data from the Purple Mountain Observatory, 
with a resolution of $\sim$53$''$. In this study, we have used higher-resolution $^{13}$CO(1--0) line data (resolution $\sim$21\rlap.{$''$}92) 
from the FUGIN survey to explore the gas distribution in the G47 cloud in greater detail. 
In this connection, we produced the integrated intensity (moment-0) map, intensity-weighted velocity (moment-1) map, 
velocity dispersion (moment-2) map, and the position-position-velocity (PPV) diagram of the $^{13}$CO(1--0) emission 
toward our selected target area.

Figure~\ref{fg5}a displays the $^{13}$CO(1--0) moment-0 map over the velocity range of [47.2, 67.2] km s$^{-1}$ toward the G47 cloud. 
No emission from the cloud associated with the N97 bubble is detected in this velocity range, suggesting that the N97 bubble is not 
physically connected to the G47 cloud. Therefore, no further analysis of the N97 bubble is included in this study. 
The moment-1 map provides insight into the velocity structure of the cloud. 
In Figure~\ref{fg5}b, the $^{13}$CO(1--0) moment-1 map of the cloud is presented, and shows significant velocity variations from 61 to 53 km s$^{-1}$. 
Figure~\ref{fg5}c exhibits the moment-2 map, which shows a localized increase in velocity dispersion, reaching up to 3 km s$^{-1}$, 
in the region between the B2 and N98 bubbles. We also derived the position-velocity (PV) diagram along the filament, shown in Figure~\ref{afg1}b, which confirms 
significant velocity variations from 61 to 53 km s$^{-1}$.
\subsubsection{Distribution of molecular gas and embedded protostars}
\label{ssmolgas}
The presence of embedded YSOs in molecular clouds serves as a key indicator of ongoing star formation activity. 
YSOs can be classified into different evolutionary stages---Class~I, II, and III---based on their infrared excess 
and the slopes of their SEDs \citep{Lada_1984, Hartmann_2005, Evans_2009}. 
Class~I YSOs are protostars surrounded by an infalling dusty envelope, leading to a higher infrared excess and strong NIR extinction. 
In contrast, Class~II YSOs have only a circumstellar disk, resulting in a lower infrared excess. 
Class~III YSOs lack circumstellar material and therefore exhibit no excess. Based on their infrared excess, YSOs occupy distinct positions 
in infrared color-color and color-magnitude diagrams, which are commonly used to identify their evolutionary stages. 
Further details about YSOs can be found in \citet{stahler04}.

In this work, we have employed the {\it Spitzer} [3.6]$-$[4.5] vs [4.5]$-$[5.8] color-color scheme to identify Class~I YSOs. 
The 3.6, 4.5, and 5.8 $\mu$m photometric magnitudes were obtained from \textit{Spitzer} GLIMPSE-I Spring\textquoteright 07 catalog \citep{Spitzer_2009}. 
In this analysis, we have only used sources with photometric magnitude errors less than 0.2 in the \textit{Spitzer} 3.6--5.8 $\mu$m bands. 
To identify Class~I sources, we have used the selection criteria [3.6]$-$[4.5] $\ge$ 0.7 mag and [4.5]$-$[5.8] $\ge$ 0.7 mag, as given by \citet[][]{Hartmann_2005,Getman_2007}. Using this scheme, we have identified a total of 34 Class~I YSOs toward the G47 cloud.

\begin{figure*}
\includegraphics[width=19cm]{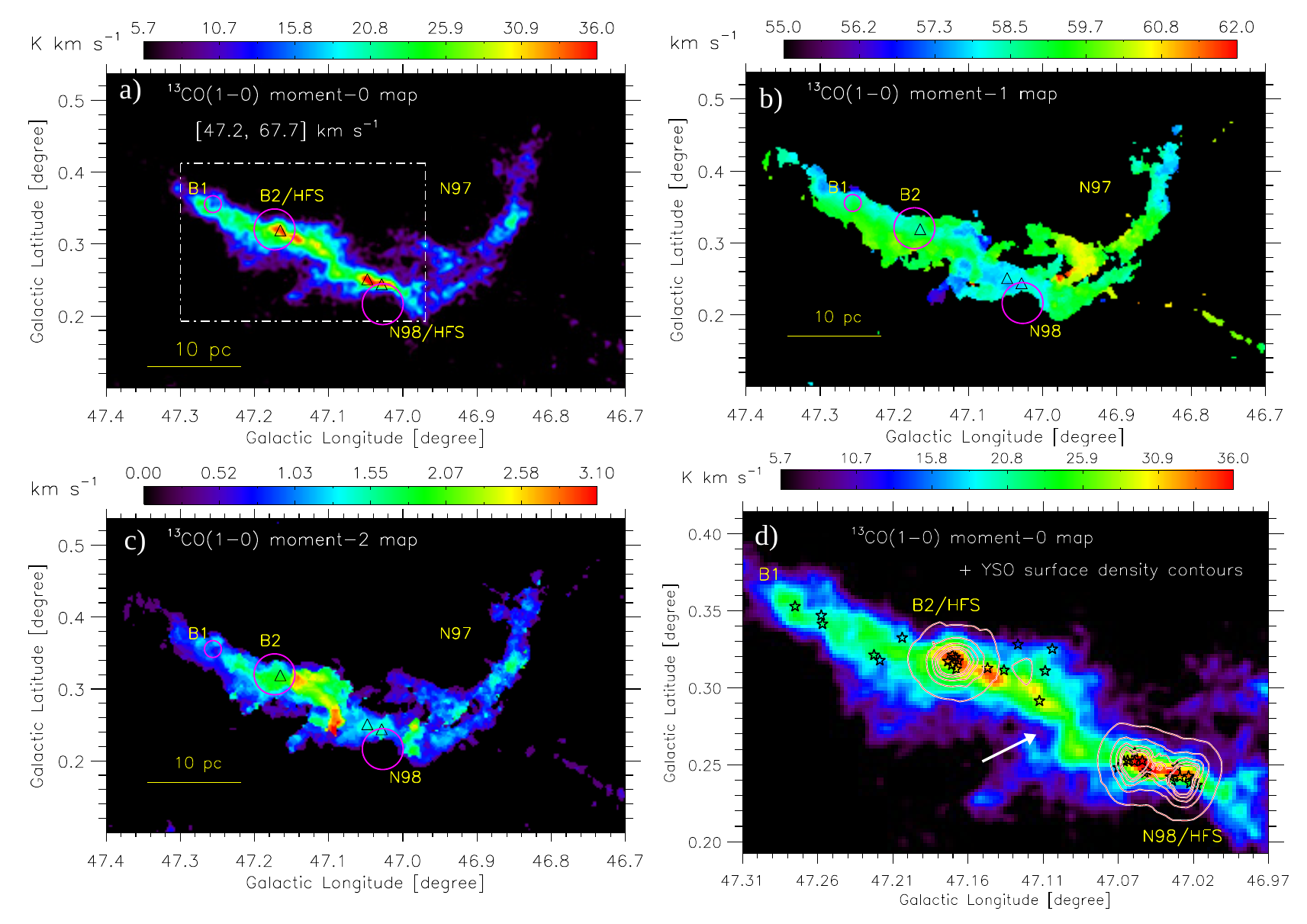}
\caption{a) FUGIN $^{13}$CO(1--0) integrated intensity (moment-0) map. 
b) FUGIN $^{13}$CO(1--0) intensity-weighted velocity (moment-1) map. 
c) FUGIN $^{13}$CO(1--0) velocity dispersion (moment-2) map. d) Overlay of YSO surface density contours (in red) on the $^{13}$CO(1--0) moment-0 map toward an area highlighted by the dot-dashed box in Figure~\ref{fg5}a. 
The surface density contours are shown with the levels of 1, 2, 3, 4, and 5 YSOs pc$^{- 2}$. The positions of Class-I and Flat-spectrum YSOs are indicated by black stars. 
In each panel, the scale bar and symbols are same as Figure~\ref{fg1}a.}
\label{fg5}
\end{figure*}

To broaden our YSO sample, we have also utilized the {\it Spitzer} [3.6] vs [3.6]$-$[24] color-magnitude scheme to identify Class~I, Flat-spectrum, Class~II, and Class~III YSOs. We have used the selection criteria, [3.6]$-$[24] $>$ 6.7 mag for Class~I YSOs and 5.45 mag $<$ [3.6]$-$[24] $<$ 6.7 mag for Flat-spectrum YSOs which is adopted from the previous work of \citet{Guieu_2010}. A total of 10 Class~I YSOs and 3 Flat-spectrum YSOs were identified using this scheme. 
A final catalogue of Class~I YSOs is generated by merging the two catalogues and counting common Class~I YSOs only once. Additionally, we 
have also included the flat-spectrum YSOs identified from [3.6] vs [3.6]$-$[24] color-magnitude scheme in our YSO sample. 
Figure~\ref{fg5}d presents the overlay of the positions of the selected Class-I and Flat-spectrum YSOs (marked by black stars) 
on the $^{13}$CO(1--0) moment-0 map, within the area highlighted by the dot-dashed box in Figure~\ref{fg5}a. 

To identify groups/clusters of stars, we applied the nearest neighbour (NN) technique 
to our complete YSO sample \citep[e.g.,][]{1985ApJ...298...80C, Gutermuth_2009,Bhadari_2020}. 
This technique was used to generate a YSO surface density map of YSOs distributed in our selected target area. 
The target area is divided into a grid of 100 $\times$ 100, corresponding to separations of 11\rlap.{$''$}41 and 6\rlap.{$''$}07 
along the $l$ and $b$, respectively. 
We adopted a distance of 4.44 kpc and NN = 6 to generate the YSO surface density map. 
In Figure~\ref{fg5}d, we also display the YSO surface density contours (in red) overlaid on the $^{13}$CO(1--0) moment-0 map. 
These contours are shown with the levels of 1--5 YSOs pc$^{-2}$. 
A total of three groups (or clusters) are found, which are spatially consistent with the locations of the ATLASGAL clumps. 
Additionally, the detection of radio continuum emission favours the presence of an OB 
star toward the N98, B1, and B2 bubbles (see Figure~\ref{fg1}b), where warm dust emission ($\sim$17--20 K; see Figure~\ref{fg2}b) is evident.
\subsubsection{Spectral Decomposition of the FUGIN $^{13}$CO(1--0) line data}
\label{sp_dec}
To gain deeper insight into the complex gas kinematics toward the G47 cloud, we performed a spectral decomposition of the $^{13}$CO(1--0) line data using the Python-based tool \texttt{SCOUSEPY} \citep{Henshaw19,Henshaw2016}. This tool allows for pixel-by-pixel fitting of multiple Gaussian components in the molecular line spectra \citep[e.g.,][]{Dewangan_2024a}. 

\begin{figure*}
\centering
\includegraphics[width=\textwidth]{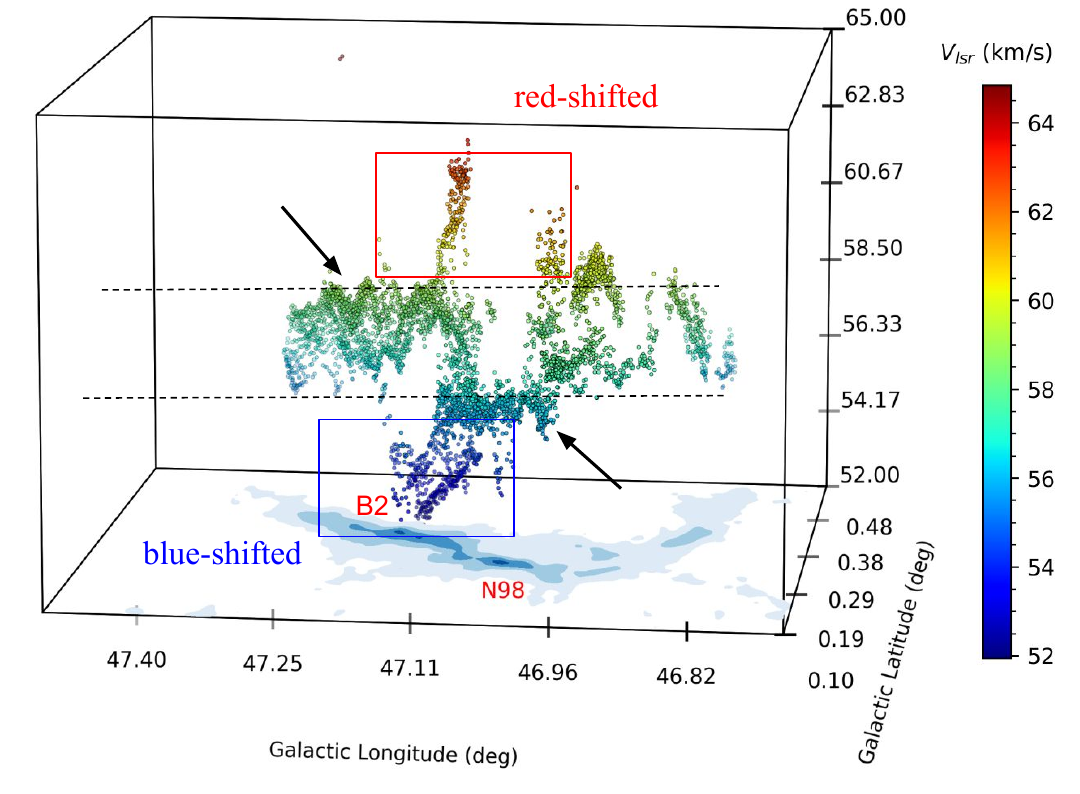}
\caption{a) PPV diagram generated after performing the spectral decomposition of the FUGIN $^{13}$CO(1--0) line data.}
\label{fg6}
\end{figure*}

To proceed, we first defined the size of the `Spectral Averaging Areas' (SAA) in pixels, which has been 
chosen as 3 $\times$ 3 pixel$^2$ for this study. The equivalent size of the SAA is 25\rlap.{$''$}5 $\times$  25\rlap.{$''$}5 (plate scale $\sim$8\rlap.{$''$}5), which is 0.54 $\times$ 0.54 pc at a distance of 4.44 kpc.
The SAAs are spatially distributed to cover all emission above a particular threshold or emission level ($\sim$3 K). 
An averaged spectrum is extracted from each SAA and fitted with single or multiple Gaussian components. 
The best-fitted parameters from the SAA averaged spectrum are then used to fit the spectra at each individual pixel. 
To better understand the resulting output, we plotted the centroid velocity of the fitted Gaussian(s) at each pixel in PPV space. 
The resulting PPV diagram is presented in Figure~\ref{fg6}. The $l$--$b$ plane of the PPV diagram displays the moment-0 map of the G47 cloud, highlighting the location of the N98 and B2 bubbles. The diagram reveals velocity oscillations between 57 and 59 km s$^{-1}$, highlighted 
the areas by two black dashed lines in Figure~\ref{fg6}. Additionally, the red-shifted emission at [61, 65] km s$^{-1}$ and the blue-shifted emission at [51, 53.7] km s$^{-1}$ are observed toward the northeast of the bubble. These red-shifted and blue-shifted velocity emissions are likely responsible for the high-velocity dispersions observed in the G47 cloud, as shown in the moment-2 map  (Figure~\ref{fg5}c). These red-shifted and blue-shifted emissions are probably caused by the expansion of the N98 and B2 bubbles, respectively.
\subsection{B-field morphology toward G47 cloud}
\label{B-field}
\begin{figure*}
\centering
\includegraphics[width=15cm]{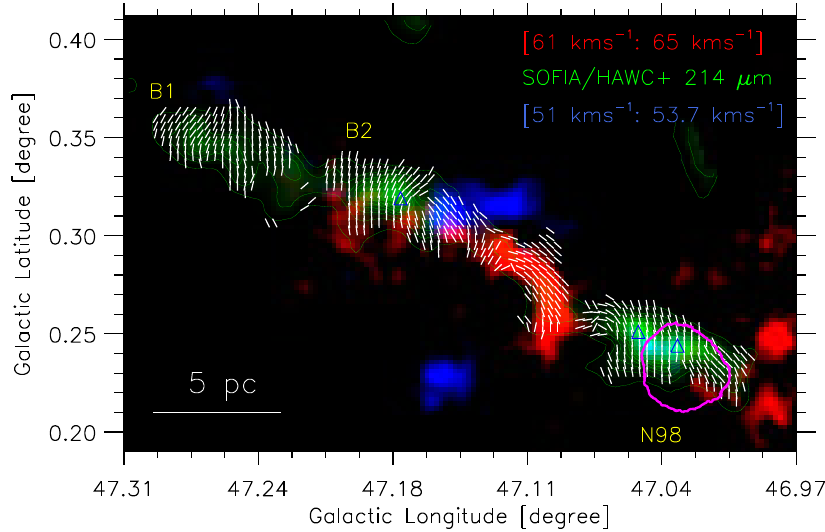}
\caption{Three-color composite map (red: $^{13}$CO(1--0) integrated emission at [61, 65] km s$^{-1}$, green: SOFIA/HAWC+ 214 $\mu$m continuum map, blue: $^{13}$CO(1--0) integrated emission at [51, 53.7] km s$^{-1}$) overlaid with segments (in white) showing the B-field orientations derived using  SOFIA/HAWC+ 214 $\mu$m toward the G47 cloud (see text for details). The SOFIA/HAWC+ 214 $\mu$m emission contours (in green) are 0.1, 0.19, 0.28, 0.37, 0.47, 0.56, 0.65, 0.75, 0.84, and 0.93 Jy beam$^{-1}$. The magenta contour highlights the SMGPS 1.3 GHz radio continuum emission (at 0.4 mJy beam$^{-1}$) associated with the N98 bubble. The scale bar and symbols are same as Figure~\ref{fg2}b.}
\label{fg7}
\end{figure*}

%
To investigate the impact of the expansion of the N98 and B2 bubbles, we analyzed the B-field morphology toward the G47 cloud. 
\citet{Stephens_2022} studied the B-field in the G47 cloud using SOFIA/HAWC+ 214 $\mu$m polarization data, 
showing that the B-field lines align with regions of high $N(\rm H_2)$, while in other regions, the field appears parallel or curved. 
The B-field morphology is generally perpendicular to high $N(\rm H_2)$ ($>$ 5 $\times$ 10$^{21}$ cm$^{-2}$) filaments, 
whereas in low $N(\rm H_2)$ ($<$ 5 $\times$ 10$^{21}$ cm$^{-2}$) filaments, the field orientations can be parallel or more randomly distributed \citep{Soler_2013}.
In low-density regions, the B-field morphology can be affected by the local gas kinematics since the field lines are effectively frozen 
into the gas. The change in the local gas kinematics can be caused by gravitational collapse, 
stellar feedback (such as expanding H\,{\sc ii} regions), or supernovae.
To investigate whether the expansion of the N98 and the B2 bubbles is responsible for the curved B-field morphology, 
we revisited the same SOFIA/HAWC+ 214 $\mu$m data to understand the observed B-field morphology toward G47 cloud. 
For better visualization, the original polarization data were binned by averaging every 25 pixels (in a 5 $\times$ 5 pixel grid) into a single pixel, resulting in a pixel scale of 22\rlap.{$''$}75. Following this, we applied the selection criteria p/$\sigma_{\rm p} >$ 3, I/$\sigma_{\rm I} >$ 100, and p $<$ 30
 to retain the physically relevant vectors.
 We integrated the red-shifted emission over the range of [61, 65] km s$^{-1}$ 
and the blue-shifted emission over the range of [51, 53.7] km s$^{-1}$, then overlaid them on the B-field orientations. 
The resulting plot is shown in Figure~\ref{fg7}. 
The curved B-field morphology is spatially coincident with the red-shifted velocity emission, while the blue-shifted velocity 
emission also overlaps with some parallel B-field orientations (see Figure~\ref{fg7}).
The PPV map shows that the gas near the N98 bubble has a lower velocity, whereas the gas near the B2 bubble moves at a 
different velocity (see arrows in Figure~\ref{fg6}). These two distinct gas flows converge at the junction, 
where both red-shifted and blue-shifted velocities are detected, indicating gas expansion. 
This pattern is also evident in the PV diagram shown in Figure~\ref{afg1}. Additionally, the filament appears to bend 
at the junction (see arrow in Figure~\ref{fg5}d), where the two flows converge.
These results suggest that the expansion of the N98 and the B2 bubbles may be responsible for the curved B-field morphology in the G47 filamentary cloud.
\subsection{Local gravity and comparison with the B-field}
\begin{figure*}
\centering
\includegraphics[width=15 cm]{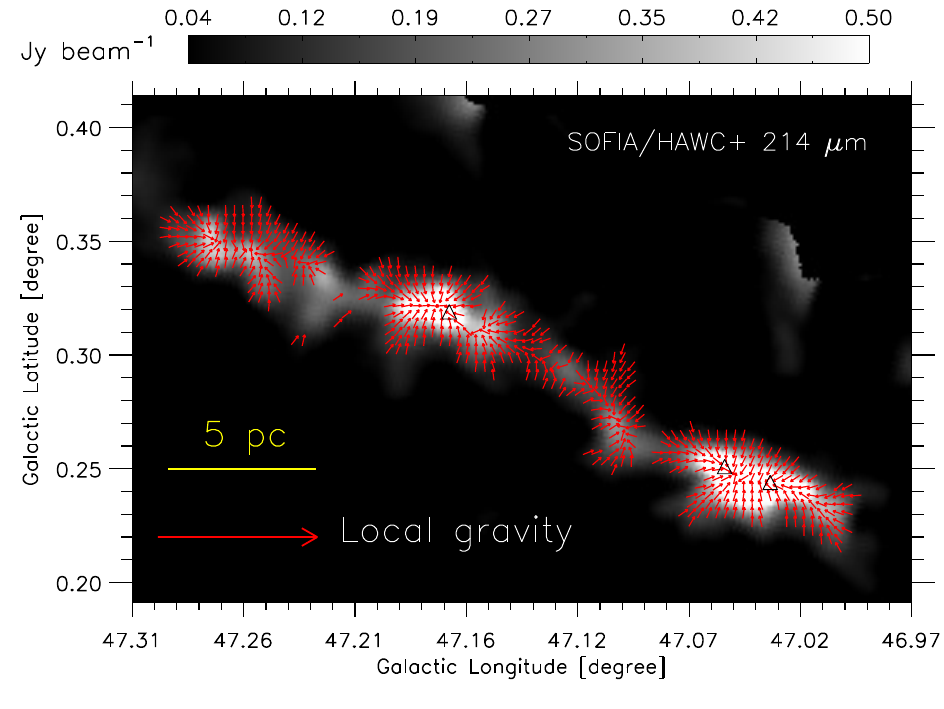}
\caption{a) Local gravity vectors (in red) overlaid on the SOFIA/HAWC+ 214 $\mu$m intensity map of the G47 cloud. The scale bar and symbols are same as Figure~\ref{fg2}b.}
\label{fg8}
\end{figure*}

To determine the local gravity vectors toward the G47 cloud, we followed the procedure outlined by 
\citet{Koch_2012a,Koch_2012b}. We used the SOFIA/HAWC+ 214 $\mu$m intensity map to compute the gravitational 
force at each pixel, considering contributions from all other pixels in the map. Only pixels with values above the 3$\sigma$ noise level are used in the calculations.
The net gravitational force at each pixel is calculated using 
\begin{equation}
\vec{F_{G,i}} = kI_i\sum_{j=1}^{n}\frac{I_j}{r_{i,j}^2}\hat{r},
\end{equation}
where $k$ is the gravitational constant (set to 1), $I_i$ and $I_j$ are the intensities at the {\it i}$^{th}$ and {\it j}$^{th}$ 
pixels, and $r_{i,j}$ is the projected distance between these two pixels. 
We assumed the gravitational constant to be 1 to simplify the calculations, as our primary focus is on the direction of the gravity vectors rather than their absolute magnitudes.
For a direct comparison with the B-field orientations, we applied the same binning and selection criteria on the gravity vectors as those used for the B-field (see Section~\ref{B-field}). 
The resulting local gravity map is presented in Figure~\ref{fg8}. The map reveals several converging centers along the crest of the filament.

\begin{figure*}
\centering
\includegraphics[width=15 cm]{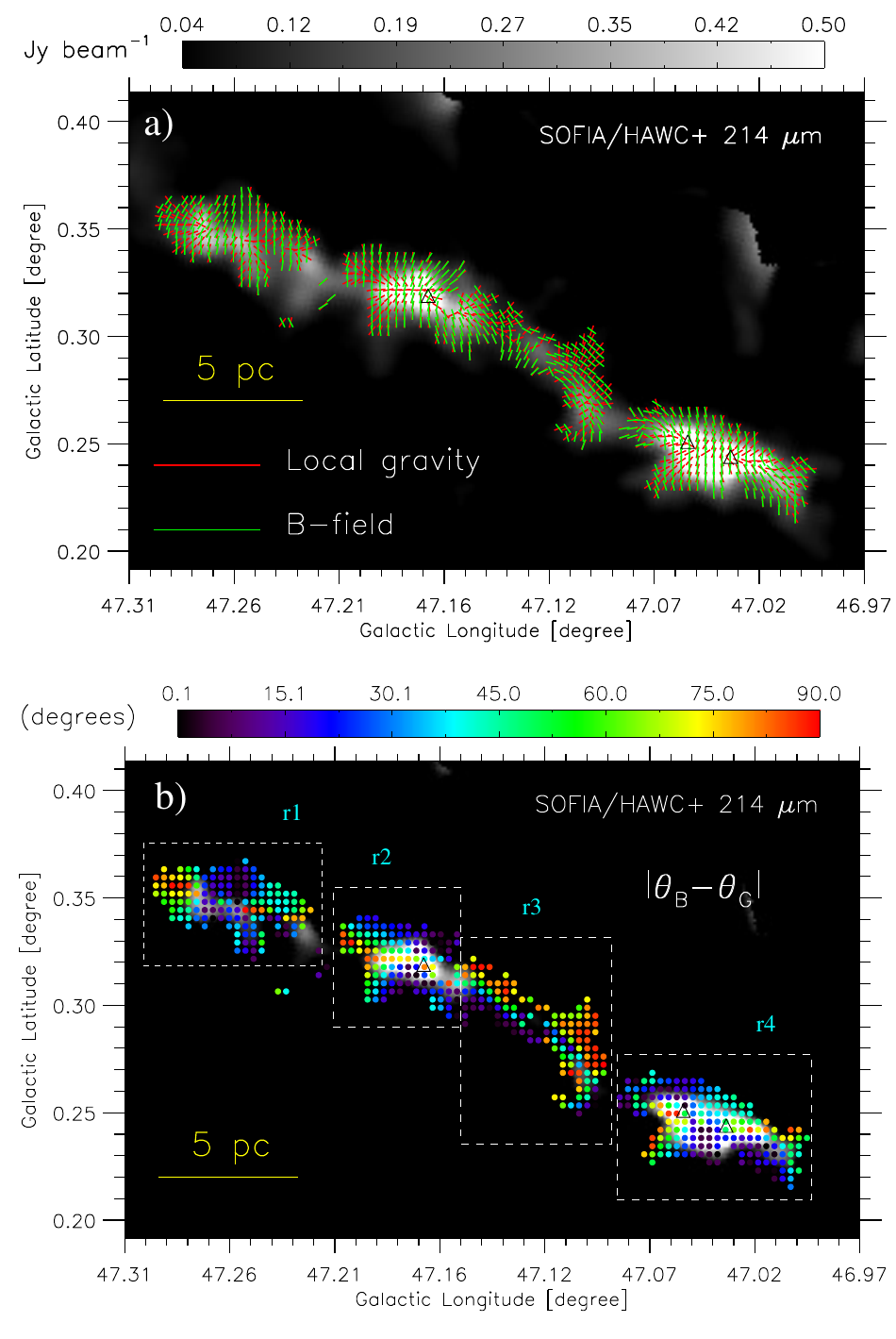}
\caption{a) Overlay of segments showing B-field orientations (in red) and local gravity vectors (in green).
b) The panel displays the difference in the position angles between B-field and local gravity (i.e., $|\theta_{\rm B} - \theta_{\rm G}|$), which is shown as color gradient ranging from 0 to 90 degrees. In both the panels, the background image is the SOFIA/HAWC+ 214 $\mu$m intensity map. In each panel, the scale bar and symbols are same as in Figure~\ref{fg2}b.}
\label{fg9}
\end{figure*}

Figure~\ref{fg9}a displays the overlay of the B-field orientation segments on the local gravity vectors, 
while Figure~\ref{fg9}b shows the spatial distribution of their relative orientations. 
This analysis provides insight into the corelation between the B-field morphology and local gravity. 

As shown in Figure~\ref{fg9}b, the B-field and local gravity tend to be more strongly aligned perpendicular to the filament’s spine in 
regions r1, r2, and r3. In contrast, their alignment is weaker along the filament’s spine. 
Figure~\ref{fg10}a highlights this pattern, showing a higher concentration of points 
at angles under 
45$^\circ$. However, in region r3, the B-field and local gravity become 
nearly perpendicular to each other, causing the histogram to peak around 90$^\circ$. 
\subsection{Energy Budget Calculations}
We have determined the contributions of gravitational ($E_{\rm G}$), kinetic ($E_{\rm T}$), and magnetic ($E_{\rm B}$) energies in the filament. 
These calculations are restricted to the region outlined by the white dashed box in Figure~\ref{fg5}a, as polarization data are unavailable beyond this area. Assuming a cylindrical geometry for the filament, the gravitational energy \citep[see][]{Fiege_2000} is given by 
%
\begin{equation}
E_{\rm G}=-GM^2/L,
\end{equation}
where $M$ (1.5 $\times$ 10$^4$ $M_{\odot}$) and $L$ ($\sim$26.1 pc) represent the filament's mass and length, respectively, and $G$ is the universal gravitational constant. 
The mass of the filament is calculated using the FUGIN $^{13}$CO(1--0) molecular line data, following the steps outlined in \citet{Xu_2018}. Since this was a repeated calculation for the same region, the detailed steps are not mentioned in this paper. The length of the filament is taken as the apparent extent along the longer axis for the region highlighted with a dot-dashed box in Figure~\ref{fg5}a.
The kinetic energy \citep[see][]{Mckee_2007} is given by 
\begin{equation}
E_{\rm T}=M\sigma_{\rm tot}^2,
\end{equation}
where $\sigma_{\rm tot}$ is the observed total velocity dispersion. The velocity dispersion is estimated using the $^{13}$CO(1--0) line data. The magnetic energy is calculated as
\begin{equation}
E_{\rm B}=\frac{1}{2}M \nu_{\rm A}^2,
\end{equation}
where $\nu_\mathrm{A}$ is the alfv{\'e}nic velocity, estimated using the DCF method \citep{Davis_1951,1953ApJ...118..113C}, defined as $\nu_{\rm A}$ = $\mathcal{Q} \times \sigma_{\rm V}/\sigma_{\theta}$.  Here,  $\mathcal{Q}$ is the correction factor (taken as 0.5), $\sigma_{\rm V}$ is the non-thermal velocity dispersion, and $\sigma_{\theta}$ is the the polarization dispersion in radians (hereafter converted to degrees). 
This equation assumes that the density and alfv{\'e}n speed are constant across the filament.
We adopted the steps used in \citet{Ngoc_2023} to create the polarization angle dispersion map. A 5 $\times$ 5 pixel$^2$ box centered on pixel {\it i} is used,  and the polarization angle dispersion is calculated as,  $\sigma_{\rm \theta} = \sqrt{\sum_{i=1}^N(\theta_i-\bar{\theta})^2/N}$. Here, $N$ is the number of pixels within the box 
and $\bar{\theta}$  is the mean polarization angle over those $N$ pixels. 
This resulting value of $\sigma_{\rm \theta}$  is then assigned to the central pixel {\it i}. 
This procedure is iterated over the entire polarization angle map, 
producing the polarization angle dispersion map for the G47 cloud (see Figure~\ref{afg2}). 
In this map, only pixels with a polarization angle dispersion below 25$^\circ$
are retained \citep[as suggested by][]{Ostriker_2001}. The mean $\sigma_{\theta}$ for the cloud is found to be 7\rlap.{$^\circ$}7 $\pm$ 4\rlap.{$^\circ$}8. 
\citet{Stephens_2022} derived the B-field strength toward the G47 cloud using velocity dispersion values of two molecular line tracers, $^{13}$CO(1--0) and NH$_3$(1,1), which trace different density regimes. In denser regions, their study used the velocity dispersion from NH$_3$(1,1), while velocity dispersion from $^{13}$CO(1--0) was used elsewhere in the cloud. This approach likely led to an underestimated B-field strength in the dense regions, due to which these regions were found to be magnetically supercritical. To obtain an unbiased estimate of the B-field energies, we used $^{13}$CO(1--0) molecular line data for the entire filamentary cloud in this study. The mean non-thermal velocity dispersion toward the cloud estimated using $^{13}$CO(1--0) is $0.93 \pm 0.57 \, \text{km s}^{-1}$, after substracting the thermal contribution ($\frac{kT}{m}$) of $0.075 \, \text{km s$^{-1}$}$, assuming a temperature of 20~K and a mass of 29~amu for $^{13}$CO. 

\begin{figure*}
\centering
\includegraphics[width=17 cm]{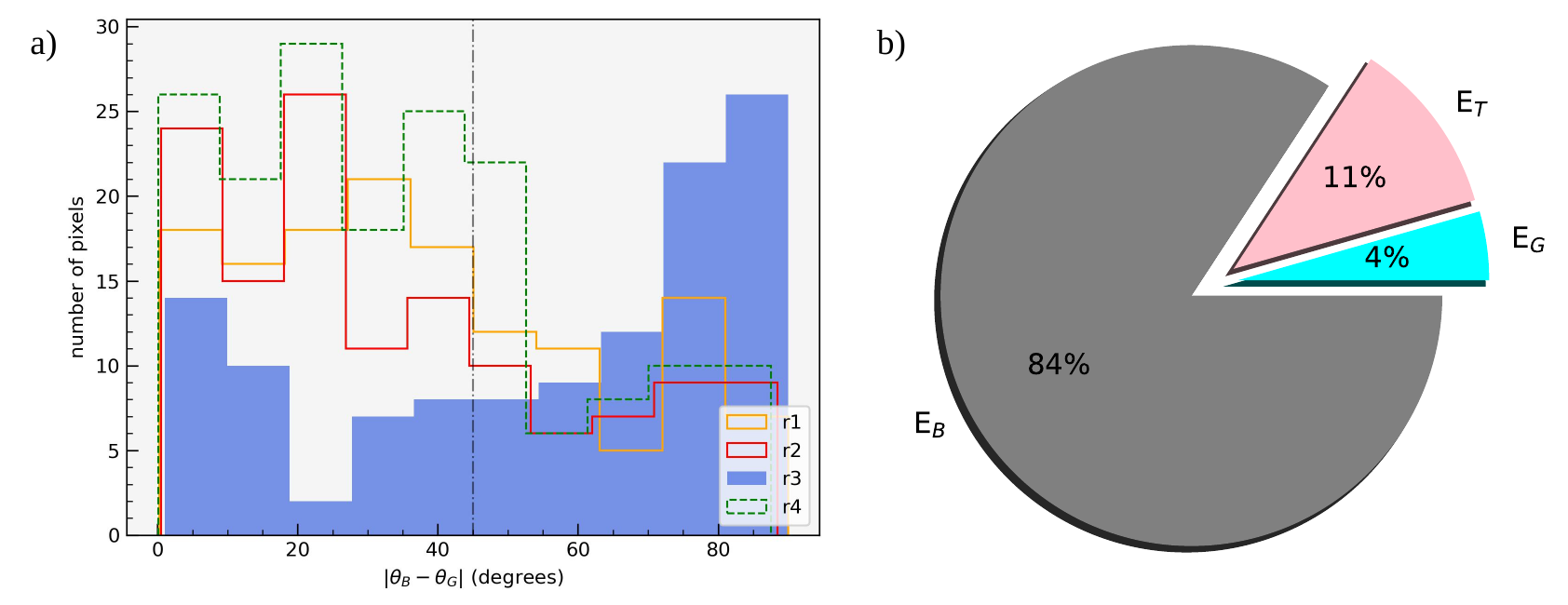}
\caption{a) The panel displays histogram plots of $|\theta_{\rm B} - \theta_{\rm G}|$ for the r1, r2, r3, and r4 regions indicated in Figure~\ref{fg9}b. 
The black vertical dot-dashed line marks the 45$^\circ$ value on the x-axis. b) The panel displays the relative importance of $E_{\rm G}$ (in cyan), $E_{\rm B}$ (in gray), and $E_{\rm T}$ (in red).}
\label{fg10}
\end{figure*}

The kinetic, gravitational, and B-field energies are listed in Table~\ref{tab3}. Figure~\ref{fg10}b illustrates the relative contributions of gravitational, kinetic, and B-field 
energies to the overall energy budget of the cloud. The B-field energy dominates 
the energy budget of the cloud ($>$ 80$\%$), followed by the kinetic and gravitational 
energies. Since the contribution of thermal motions to the kinetic energy is negligible, the total kinetic energy is considered as the turbulent energy of the cloud.
\begin{table}
\centering
\small
\renewcommand{\arraystretch}{1.2}
\setlength{\tabcolsep}{0.8 cm}
\caption{Table lists the values of different energies determined for the cloud G47.}
\label{tab3}
\begin{tabular}{ccc}
\hline
\hline
\textit{Parameter} & \textit{Value}  \\
\hline
$E_{\rm G}$&  7.27 $\times$ 10$^3$ M$_{\odot}$ m$^2$ s$^{-2}$\\
$E_{\rm T}$&  1.85 $\times$ 10$^4$ M$_{\odot}$ m$^2$ s$^{-2}$\\
$E_{\rm B}$&  1.37 $\times$ 10$^5$ M$_{\odot}$ m$^2$ s$^{-2}$\\
\hline 
\end{tabular}
\end{table}

\section{Discussion}
\label{discussion}
\subsection{Hub-filament systems and formation of massive stars}
HFSs are structures where parsec-scale, low-density filaments converge into a central hub \citep{myers09}. 
These configurations are commonly observed in star-forming regions, with the central hub serving as an active site of star formation, 
including the formation of massive stars. Material is funneled from parsec-scale molecular filaments into the central hub, 
where massive stars may form \citep{Motte+2018,Kumar_2020}. This process is confirmed by the detection of longitudinal velocity gradients 
along the filaments \citep{Chen_2020b}. HFSs have been observed on small scales \citep[i.e., 0.55--6 pc;][]{Dewangan_2024a,Dewangan_2024b,Bhadari_2025} 
to large-scale \citep[$\sim$ 10--20 pc;][]{Schneider_2012,Kumar_2020,Zhou_2022,Bhadari_2022,Bhadari_2023} in star-forming regions. 
Hence, HFSs are key sites for investigating massive star formation activity. 
However, most existing studies report only a single HFS in a cloud, and examples of multiple HFSs in a single cloud are still rare in the literature \citep[e.g.,][]{Dewangan_2024a,Maity_2024}.

\begin{figure*}
\includegraphics[width=18cm]{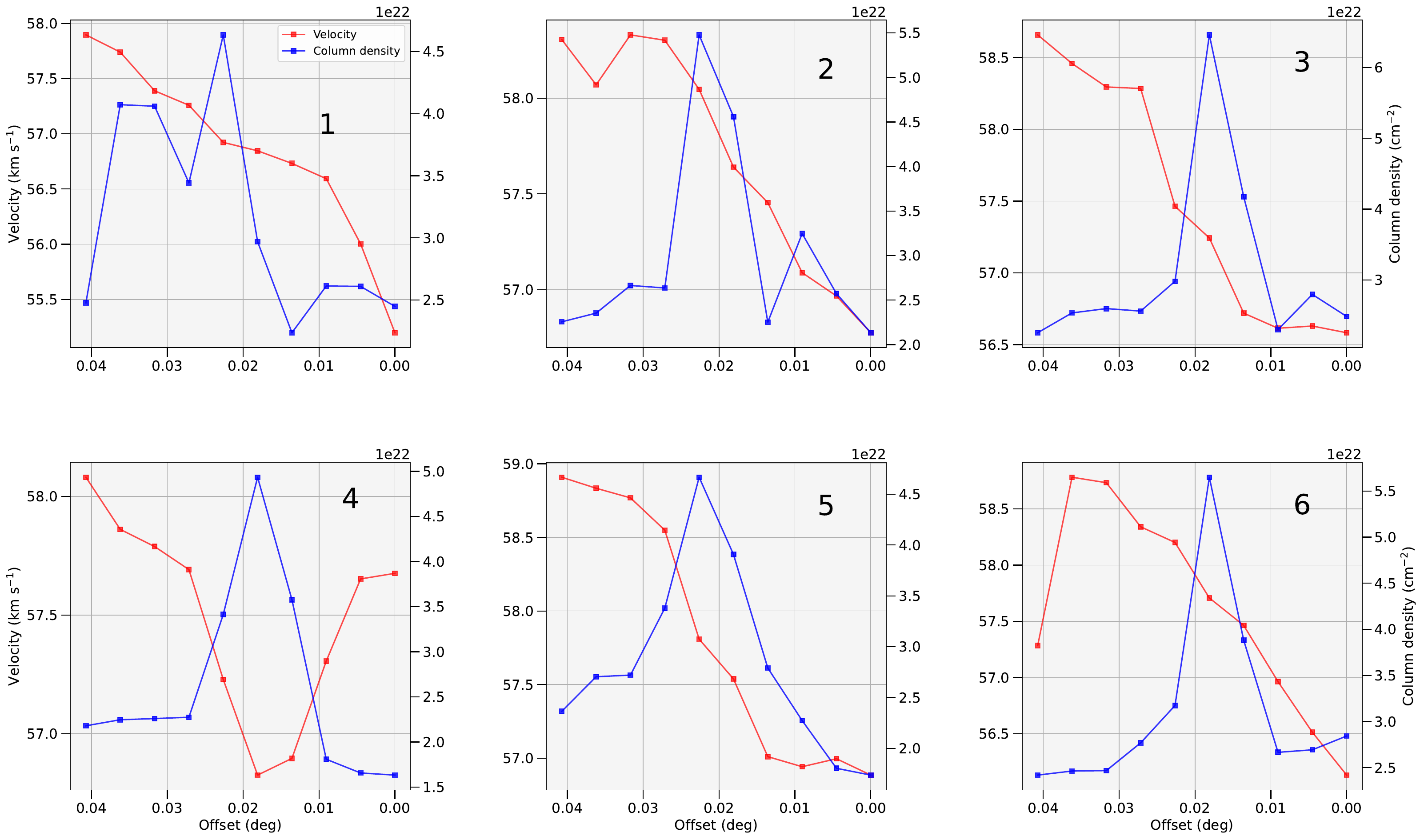}
\caption{1--6: Panels display the velocity profile (in red) and the column density profile (in blue) along six cuts indicated in Figure~\ref{fg2}a.} 
\label{fg11}
\end{figure*}

The Bolocam clumps (including ATLASGAL clumps) in the G47 cloud have been proposed as potential sites for massive star formation 
\citep[e.g.,][]{Xu_2018}. However, the formation mechanisms of massive stars in the G47 cloud 
remain poorly understood in the literature. 
In this work, the analysis of the SMGPS radio continuum emission map and the {\it Herschel} $T_{\rm d}$ map 
reveals the ionized and warm dust emissions toward the B1, B2, and N98 bubbles, confirming 
the existince of massive OB stars in each bubble. A detailed examination of multi-wavelength data has revealed the existence of a single HFS candidate associated with the B2 and N98 bubbles in the G47 cloud 
(see Section~\ref{HIRES}). 
The radial $N(\rm H_2)$ and velocity profiles show distinct HFS signatures, where the $N(\rm H_2)$ peak coincides with the minima of the velocity profile \citep[e.g.,][]{Zhou_2023}. 
To explore this further, we analyzed the velocity and $N(\rm H_2)$ profiles along the minor axis of the filament at six positions (see white lines in Figure~\ref{fg2}a). The resulting profiles are presented in Figure~\ref{fg11}. 
Notably, the distinct HFS signature is clearly visible toward the HFS associated with the B2 bubble (see panel~4 in Figure~\ref{fg11}), 
but is absent elsewhere in the filament (see Figure~\ref{fg11}). 

The global non-isotropic collapse scenario (GNIC) explains the physical process driving star-formation in HFSs \citep{Tige_2017, Motte+2018}. 
The model explains that, in a hub/ridge filament system, gas flows along the filaments into the central hub, where massive dense cores (MDCs) are formed. 
Over time, these MDCs evolve into infrared-quiet protostellar objects containing low-mass stellar embryos. These embryos then evolve into infrared-bright high-mass protostars through gravitational inflows, which eventually grow into massive stars and announce their presence through H\,{\sc ii} regions. 
The ionized gas pressure from the H\,{\sc ii} region can disrupt the parent HFS, erasing its structural signatures or initial conditions. 
As a result, identifying an HFS that has been disrupted by the feedback of the massive star becomes challenging. Considering this evolutionary sequence, the HFSs observed in the G47 cloud appear to be in a more evolved stage, influenced by feedback of massive stars in the form of H\,{\sc ii} regions. 

A semi-analytic model by \citet{whitworth21} examines the impact of ionizing feedback from an O-type star formed within a filament. The study presents two scenarios. In the first, the powerful radiation from the massive star disrupts the filament, dispersing ionized gas into the surrounding medium. In the second scenario, if the filament is sufficiently dense and/or the O star emits ionizing photons at a relatively low rate, the accretion flow onto the filament can limit the escape of ionized gas, potentially trapping the ionizing radiation \citep[see also][]{Dewangan_2022}. This trapping of ionized gas increases the turbulent energy and change the gas dynamics in the vicinity of the expanding H\,{\sc ii} region.
In the case of G47 cloud, the second scenario appears to be applicable, cosnidering the complex dynamics around the B2 and N98 H\,{\sc ii} regions.
\subsection{Impact of  H\,{\sc ii} regions on the B-field structure of the G47 cloud}
The POS B-field structure of the G47 cloud was analyzed by \citet{Stephens_2022} using SOFIA/HAWC+ 214 $\mu$m polarization data. 
Their study revealed that the B-field orientations are predominantly perpendicular to the major axis of the filament. 
However, in certain regions of the cloud, the B-field exhibits a curved morphology, the origin of which remains unclear. 
As shown in Figure~\ref{fg7}, the curved B-field morphology spatially coincides with the red-shifted emission, while the blue-shifted emission is associated with regions where the B-field is oriented parallel to the filament. Since B-field lines are effectively frozen into the molecular gas due to the high conductivity of the medium, any changes in gas dynamics directly impact the B-field structure of the cloud \citep{Mouschovias_1991}. 
Gas motions, such as compression, expansion, or turbulence can distort or realign the B-field lines \citep{Peters_2011,Liu_2022,2023ApJ...944..139T}. 
This coupling between gas dynamics and the B-field plays a crucial role in shaping the observed B-field morphology in molecular clouds \citep{Hennebelle_2019}. 
The red-shifted and blue-shifted components are located between the expanding N98 and B2 bubbles, where the surrounding gas 
exhibits different velocities (see white arrows in Figure~\ref{afg1}). The convergence of these two distinct gas flows at the interface 
between the bubbles likely drives gas expansion, producing the observed blue- and red-shifted velocity components.

In general, the B-field orientations toward the dense filaments are perpendicular to the major axis of the filament, while the field orientation is found parallel/random to the low density filaments which are connected to the main filament \citep[e.g.,][]{10.1093/mnras/stt1849,2013ApJ...774..128S, 2016A&A...586A.135P,2017A&A...603A..64S}. 
However, high resolution observations have shown that B-field orientations in the low-density regions of filamentary clouds are much more complex, influenced by local gas motions. Based on the study of B-field orientations toward IC 5146 using the JCMT BISTRO data, \citet{2019ApJ...876...42W} proposed that the parsec-scale filaments initially form under the influence of B-fields. 
Once the filaments become magnetically supercritical, they undergo local collapse, causing the B-field to bend. 
A similar behaviour in change of B-field orientations has also been observed at smaller scales ($\sim$0.01 pc) with ALMA \citep[][]{2021ApJ...915L..10S,2024arXiv240810199Z}, where the B-field orientation is dragged from a perpendicular to a parallel alignment along infalling spirals. 
The magnetohydrodynamic (MHD) simulation by \citet{10.1093/mnras/sty2018} also suggests that during gravitational contraction in a filament, local gas flows can drag the B-field direction from a perpendicular to a parallel orientation along the filament. 
This change in local gas motions can be driven by stellar feedback from massive stars. 
It has been observed that stellar feedback (i.e.,  H\,{\sc ii} regions or supernovae) can affect the B-field morphology in 
the molecular clouds \citep{Heiles_1989,Soler_2018,Tahani_2019}. 
The H\,{\sc ii} regions powered by massive OB stars can affect the kinematics of the surrounding molecular gas, 
which in turn affects the B-field structure of the parent molecular cloud \citep{2023ApJ...944..139T}. 
This effect has also been confirmed by MHD simulations \citep{Krumholz_2007, Arthur_2011}. 
Recently, \citet{Bij_2024} investigated the interaction between B-fields and stellar feedback 
in the high-mass star-forming region RCW 36 using SOFIA/HAWC+ data. 
Their analysis showed that stellar feedback played a key role in shaping the gas dynamics within RCW 36. 
The study revealed changes in the B-field morphology surrounding the expanding bipolar H\,{\sc ii} region, 
highlighting the impact of stellar feedback on the local B-field structure.

Collectively, these studies suggests that the B-field structure in molecular clouds can be affected by local gas kinematics 
driven by stellar feedback. In the G47 cloud, the expansion of the N98 and B2 bubbles likely drives the observed gas velocities, 
leading to the formation of the curved B-field morphology.
\subsection{Role of B-field in the G47 cloud}
Energy budget calculations in the G47 cloud have revealed that the B-field dominates the energetics, 
followed by turbulence and gravity. In general, the energetics in the filamentary clouds are found 
to be dominated by B-fields \citep{Arzoumanian_2021, Chung_2022, Chung_2023}. 
However, the secondary contributions of turbulence and gravity are seen to be varying accross various clouds. 
\citet{Nakamura_2008} discussed the star-formation scenario in B-field dominated clouds using MHD simulations. 
The initial conditions in these simulations are set such that the B-field is the dominant force, followed by turbulence and gravity.
In such systems, turbulence is strong enough to counteract gravitational settling along B-field lines. 
However, large-scale turbulent converging flows can overcome magnetic resistance in some regions of the cloud, 
leading to the formation of high-density regions. In these regions, gravity becomes strong enough to overcome magnetic resistance, 
ultimately driving star formation. This scenario indicates that star formation in B-field-dominated clouds is initiated 
by turbulent converging flows, with the process regulated by B-fields. The B-field lines act as channels, guiding the converging flows. 
However, gas motions are largely confined along the B-field lines due to the Lorentz force.

\citet{Chen_2020} performed MHD simulations of a self-gravitating turbulent filament embedded within a sheet-like layer, 
formed by parallel plane converging flows driven by shock compression. Their results showed that the B-field, oriented perpendicular 
to the filament's major axis, created a preferred direction for the converging gas flows, leading to a systematic velocity gradient 
along this axis. Additionally, the peak of the $N(\rm H_2$) profile in the filament was spatially coincident with the center 
of the steep velocity gradient. Such patterns in velocity and $N(\rm H_2$) profiles 
can be a potential observational indicator of an accreting filament in a sheet-like structure. 
In Figure~\ref{fg11}, the velocity and $N(\rm H_2)$ profiles along the minor axis of the filament 
at six positions are presented. These profiles resemble the accreting filament structure as proposed by \citet{Chen_2020}.
Another possible explanation for this velocity gradient is cloud rotation. However, the rotation of filamentary clouds typically 
produces distinct red-shifted and blue-shifted velocity components, which are often separated by a sharp transition in 
integrated position-velocity diagrams \citep[e.g.,][]{Kong_2018,Lobos_2019,Alvarez_2021}. However, the PV diagram of the G47 cloud 
lacks any such characteristic features (see Figure~\ref{afg1}), suggesting that the observed velocity gradient is 
unlikely to be caused by cloud rotation. We did not analyze the velocity profiles toward other regions of the cloud with high velocity dispersion, 
as the velocity structure of in those regions of the cloud is highly affected by the feedback of the N98 and B2 H\,{\sc ii} regions. 
The observed velocity and $N(\rm H_2$) profiles provides evidence that the G47 cloud is accreting material along its major axis, 
exibhiting clear signatures of converging flows.
According to \citet{Nakamura_2008}, as turbulence in a cloud decays and gravity eventually overcomes the B-field pressure, 
the filament becomes thermally supercritical. This gravitational contraction is then guided by the B-field lines. 
In the G47 cloud, we examined the alignment between the B-field and the local gravitational field direction. 
A clear alignment between the B-field and local gravity is observed perpendicular to the filament's spine (see Figure~\ref{fg9}b). 
This suggests that the G47 cloud may undergo gravitational contraction along the B-field lines once it becomes 
magnetically supercritical. However, under the current conditions, turbulence dominates over gravitational forces. 
Meanwhile, the G47 cloud has become supercritical in regions with higher $N(\rm H_2)$ ($>$10$^{22}$cm$^{-2}$), particularly within the ATLASGAL clumps. However, the current resolution of SOFIA/HAWC+ at 214 $\mu$m ($\sim$18\rlap.{$''$}2) is insufficient to resolve the B-fields within these higher $N(\rm H_2)$ regions.
Despite this limitation, the resolution is adequate for estimating the criticality of the filament on larger scales. While the filament as a whole is not magnetically supercritical, localized regions within the cloud exhibit gravitational dominance over the B-field, leading to ongoing star formation activity. To confirm this interpretation, high-resolution polarimetric observations at longer wavelengths are required. As the filament continues to evolve and becomes completely supercritical, it is expected to trigger additional episodes of star formation throughout its structure.

Overall, the B-fields are dynamically important in the G47 cloud, acting as channels for converging gas flows 
and potentially guiding gravitational contraction once the cloud reaches a magnetically supercritical state.
\section{Summary and Conclusion}\label{conclusion}

We present a multi-wavelength study of the filamentary cloud G47 (d $\sim$4.44 kpc), which hosts the MIR bubbles N98, B1, and B2. The key findings of our study are outlined below:

\begin{enumerate}

 \item{ The 1.3 GHz continuum emission from the SMGPS survey is detected toward B2 for the first time, in addition to the previously known H\,{\sc ii} regions N98 and B1.}

\item{A detailed examination of the unWISE 12.0 $\mu$m, {\it Spitzer} 8.0 $\mu$m image, along with the {\it Herschel} $N(\rm H_2$) and $T_{\rm d}$ maps, reveals two HFS candidates associated with the H\,{\sc ii} regions B2 and N98 powered by massive OB stars, which indirectly favors the GNIC scenario to explain massive star formation in the G47 cloud. }

\item{The PPV diagram of $^{13}$CO(1--0) shows significant velocity variations toward areas between B2 and N98, where the B-field morphology exhibits significant curvature, and high velocity dispersion is observed, which can be explained by the expansion of the H\,{\sc ii} regions B2 and N98. }

\item{The energy budget calculations reveal that the B-field energy dominates the overall dynamics of the cloud, accounting for more than 80$\%$ of the total energy followed by turbulence ($\sim$11$\%$) and gravity ($\sim$4$\%$).}

\item{The study of the relative orientations between the B-fields and local gravity in the G47 cloud shows that the filament may undergo gravitational contraction along the B-field lines once it becomes magnetically supercritical.}

\item{The radial $N(\rm H_2$) and the radial velocity profiles of G47 cloud show the signatures of converging gas flows in a sheet-like structure as proposed by \citet{Chen_2020}.}

Overall, this study reveals how magnetic fields, gas dynamics, and stellar feedback intricately shape star formation in G47.

\end{enumerate}

\section{Acknowledgments}
We thank the anonymous referee for carefully reviewing the manuscript and providing valuable comments and suggestions, which helped improve the scientific content of this paper. The research work at Physical Research Laboratory is funded by the Department of Space, Government of India. We acknowledge the support from Ajman University through the Internal Research Grant No. DRGS Ref. 2024-IRG-HBS-7. This research was also supported by King Faisal University, Al-Ahsaa, Saudi Arabia, under Proposal Number KFU242705. PS was partially supported by a Grant-in-Aid for Scientific Research (KAKENHI Number JP22H01271 and JP23H01221) of JSPS. The MeerKAT telescope is operated by the South African Radio Astronomy Observatory, which is a facility of the National Research Foundation, an agency of the Department of Science and Innovation. The National Radio Astronomy Observatory is a facility of the National Science Foundation operated under cooperative agreement by Associated Universities, Inc. This publication makes use of data from FUGIN, FOREST Unbiased Galactic plane Imaging survey with the Nobeyama 45 m telescope, a legacy project in the Nobeyama 45 m radio telescope. This research is based on observations made with the NASA/DLR Stratospheric Observatory for Infrared Astronomy (SOFIA). SOFIA is jointly operated by the Universities Space Research Association, Inc. (USRA), under NASA contract NNA17BF53C, and the Deutsches SOFIA Institut (DSI) under DLR contract 50 OK 0901 to the University of Stuttgart. This work is based, in part, on observations made with the \textit{Spitzer} Space Telescope, which is operated by the Jet Propulsion Laboratory, California Institute of Technology under a contract with NASA. This research made use of Astropy, a community-developed core Python package for Astronomy \citep{astropy:2013,astropy:2018,astropy:2022}.





\bibliographystyle{aasjournal}
\bibliography{bibfile}{}

\appendix
\restartappendixnumbering
\newpage
\section{PV diagrams}
\begin{figure*}[htbp!]
\includegraphics[width=18cm]{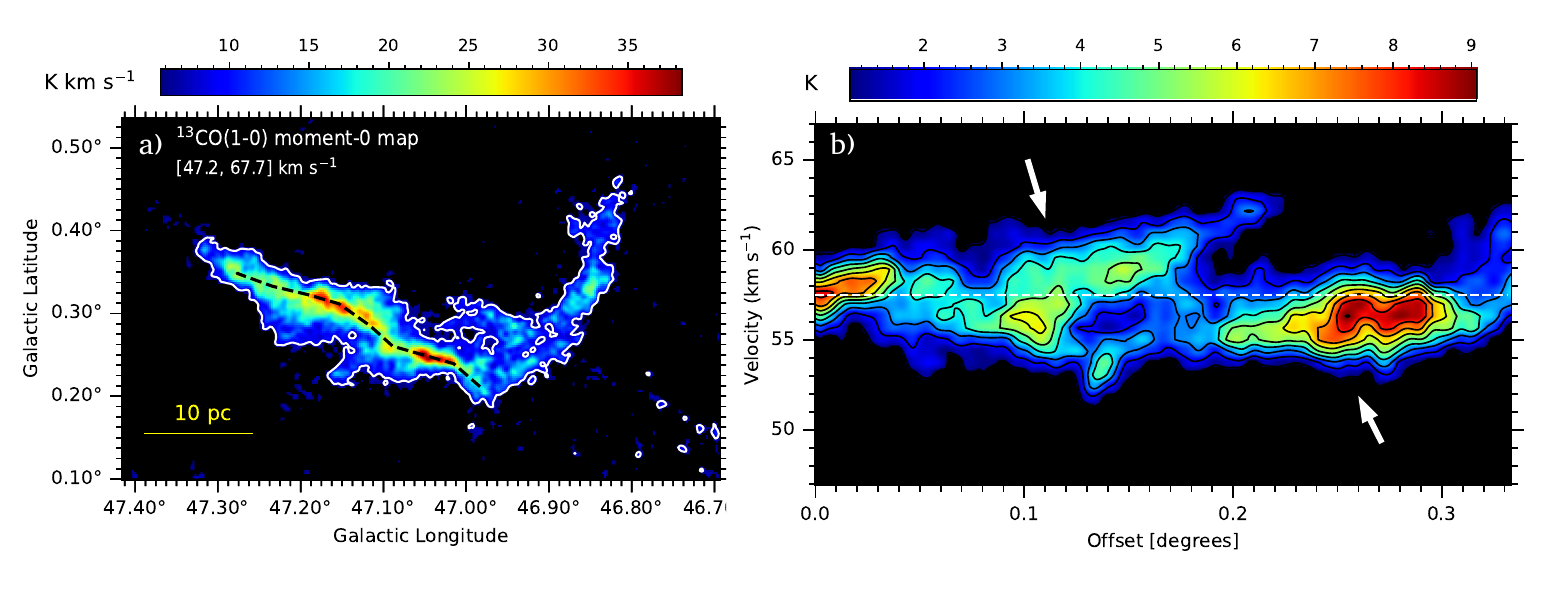}
\caption{a) $^{13}$CO(1--0) integrated intensity (moment-0) map. The white contour level is at 4$\sigma$ (1$\sigma$ $\sim$1.88 K km s$^{-1}$). The black dashed curve represents the path along which the PV diagram is generated. b) The PV diagram is derived along the black curve shown in Figure~\ref{afg1}. The offset at 0 degrees corresponds to the northeast end of the black curve. The black contours represent intensity levels, starting from 1$\sigma$ ($\sim$1.05 K) and increasing to 15 K linearly. The white dashed line represents the systemic velocity of the cloud at $\sim$57.5 km s$^{-1}$.} 
\label{afg1}
\end{figure*}

We generated the position-velocity (PV) diagrams along the curve shown in Figure~\ref{afg1}a. The resulting PV diagram is shown in Figure~\ref{afg1}b.

\section{Polarization dispersion map}

\begin{figure}
\includegraphics[width=13 cm]{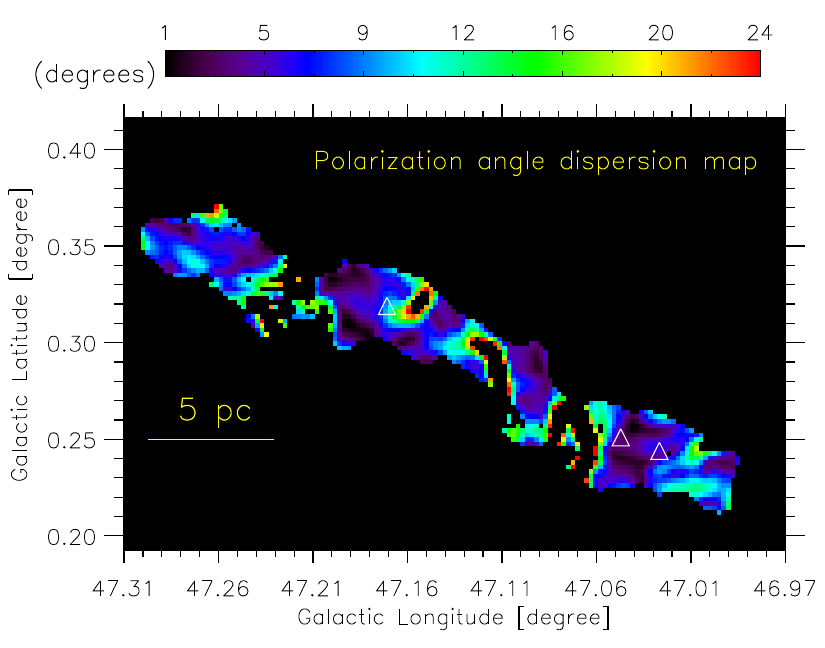}
\caption{Polarization angle dispersion map.}
\label{afg2}
\end{figure}

\end{document}